# (Mis)align: A Simple Dynamic Framework for Modeling Interpersonal Coordination

Grace Qiyuan Miao[1,*], Rick Dale[1], Alexia Galati[2]

[1]Department of Communication, University of California, Los Angeles

[2]Department of Psychological Science, University of Carolina, Charlotte

*Corresponding author: q.miao@ucla.edu

## Author Note

Grace Qiyuan Miao  https://orcid.org/0000-0001-7171-0542

Rick Dale  https://orcid.org/0000-0001-7865-474X

Alexia Galati  https://orcid.org/0000-0001-7593-6629

The authors declare that there are no conflicts of interest with respect to this manuscript.

This material is based upon work supported by the National Science Foundation under Grant #2120932 *Identifying multimodal signatures of coordination to understand joint performance in diverse tasks*.

Correspondence should be addressed to Grace Qiyuan Miao, University of California, Los Angeles, Department of Communication, 2133 Rolfe Hall, Los Angeles, CA 90095, United States. Email: q.miao@ucla.edu.

**Abstract**

As people coordinate in daily interactions, they engage in different patterns of behavior to achieve successful outcomes. This includes both synchrony – the temporal coordination of the same behaviors at the same time – and complementarity – the coordination of the same or different behaviors that may occur at different relative times. Using computational methods, we develop a simple framework to describe the interpersonal dynamics of behavioral synchrony and complementarity over time, and explore their task-dependence. A key feature of this framework is the inclusion of a task context that mediates interactions, and consists of active, inactive, and inhibitory constraints on communication. Initial simulation results show that these task constraints can be a robust predictor of simulated agents' behaviors over time. We also show that the framework can reproduce some general patterns observed in human interaction data. We describe preliminary theoretical implications from these results, and relate them to broader proposals of synergistic self-organization in communication.

**(Mis)align: A Simple Dynamic Framework for Modeling Interpersonal Coordination**

People coordinate with each other to perform joint tasks in a wide variety of circumstances. Such coordination involves individuals performing synchronized and complementary behaviors to achieve successful outcomes. Researchers of interpersonal coordination have used a variety of terms to describe patterns associated with behavioral synchrony: alignment [1], entrainment [2], mimicry [3], accommodation [4], and contagion [5], among other terms (see [6] for a systematic review). These terms capture phenomena where individuals exhibit the same behavior at about the same time. Beyond this interpersonal synchrony, researchers have identified other coordination patterns that allow for more complementarity in how individuals organize their behavior over time. Terms like synergy, coupling, misalignment, or complementarity highlight that individuals can coordinate without exhibiting the identical behaviors [7,8,9,10]. Indeed, these two patterns can be accommodated in many everyday collaborative situations. For example, when searching for a missing item in a room, two people could both look at the bed (i.e. synchronized joint attention), or one at the bed while the other at the table (i.e. complementary visual search). These behavioral patterns are intuitive and pervasive. How do people balance synchronized and complementary behaviors during collaboration, and what are the underlying cognitive processes that make them possible? These questions pertain to theories of social interaction and have implications for how we optimize collaboration and task performance [11,12,13].

One possibility is that synchrony and complementarity are supported by cognitive control processes. Consider two interlocutors engaging in friendly turn-taking, where one speaks and the other listens (i.e. complementarity). Inhibition in this case (e.g., when one partner suppresses what they want to say until the other finishes) could be important for minimizing the interference of information that would not advance the joint task goals. Indeed, there is evidence that, in situations involving multiple perspectives and potential ambiguity, language users need to actively inhibit their own perspective to process language appropriately [14]. These cognitive functions, central to interpersonal synchrony, could also be used to identify leader-follower roles in human communication [15], where the leader initiates more actions and the follower might inhibit actions to allow smooth turn-taking and consequently successful collaboration.

One way to investigate these cognitive questions is through computational simulation [16,17]. Computational models represent a simplification of a real-world system, allowing researchers to test experimental treatments or what-if scenarios that are practically infeasible in real-life settings. Moreover, simulated data complement empirical data collected in laboratories, granting researchers greater power to draw theoretical implications ([16,18,19,20] for discussions of how modeling supports theory development).

One open question is *how* humans determine when to synchronize with or complement each other, particularly in tasks that permit both options. There is ongoing debate about the prevalence of alignment in conversation [21], its reliance on priming as a mechanism [1,22], the default status of egocentric processes during alignment [23,24,25,26,27], and its benefits to joint performance. Some studies show negative effects of excessive alignment on joint performance [28,29,30], others highlight its importance [31,32,33], and yet others suggest that alignment is highly dynamic and complex, so its benefits are dependent on context [34]. Although computational models have been developed for joint action, most are related to motor control [35], and few account for the emergence of partners' behavior in conversation (see [36] for review). Proposals of computational models of coordination in conversation are often schematic (e.g., [37,38]), and have not yet been implemented and tested [20,39].

In the present work, we develop a simplified framework to describe dynamics of synchrony and complementarity in dyads and explore their task-dependence. The framework is deliberately simple, using coupled linear equations that could implement behavioral dynamics akin to coupled oscillators [37] yet accounts for additional interactional features. This permits potentially integrating hypotheses about cognitive and social constraints. In the version presented here, we show that this modeling framework could offer insight into the emergence of synchrony and complementarity, especially by including parameters related to the task and context of interaction. In the next section, we summarize our modeling approach, and compare it to prior computational strategies. Following this, we give a formal specification of the model, consisting of two agents coupled by an interactive context. We then simulate thousands of "conversations" by testing all possible parameters of the model, and discuss their theoretical implications.

## Modeling Approach

Recently, computational investigation of human communication and other behavior has often used unsupervised machine learning and related methods [40]. The power of these methods derives from their ability to take in huge amounts of data and train a model from scratch. These methods produce models that can sometimes generate surprisingly human-like behavior [41]. Despite these strengths, the inner workings of many-layered deep learning neural networks can be difficult to interpret [42].

Another approach to computational investigation is to use models that are derived from "first principles." These are models in which modeling choices are theoretically inspired so that parameters can be interpreted transparently in their development and resulting behavior. Starting from first principles, with theoretically-motivated equations, means that there are relatively fewer parameters plugged in directly to the model, and can be interpreted relatively transparently (see also [39] for a recent model on turn-taking).

Interpersonal coordination is a domain in which computational models often draw from first principles in this way, allowing for greater transparency and interpretability (see [43] for review). For example, some models that follow this approach specify dynamical systems equations, with parameters carefully chosen to represent particular aspects of interpersonal coordination. Many such models have been developed to account for joint action in terms of motor control, in tasks such as finger tapping [44,45,46] or using hand movements to shepherd target objects [35] or to avoid collision [47].

Other models of interpersonal coordination transparently capture dynamics in activities that extend beyond motor coordination, including conversation. These models typically focus on activities where agents aim to achieve synchrony or convergence. One prominent approach uses the theory of coupled oscillators as a mathematical foundation, defining a system that executes periodic behavior [48]. A key characteristic of the coupled oscillator model is that the system will always achieve synchronization eventually, whether or not the oscillators are identical [48]. Models using this approach have been developed for various contexts, including psychotherapy [49], musical coordination [45, 50, 51], and romantic couples discussing health behaviors [52]. In these types of interpersonal activities, the goal involves synchronization. For instance, music ensembles aim to play notes at synchronized beats, while romantic couples or therapist-client dyads strive to reach consensus by the end of their conversations. The nature of these activities – centered on convergence – makes coupled-oscillator models suitable for abstracting formalisms of behavior.

However, human interaction does not depend solely on synchrony [47,53]. There is a need to create models representing more diverse and generalizable patterns of human interactions, which encompass both synchrony and complementarity in various contexts. One example of this effort is a simple dynamic system of coupled agents programmed to "take turns" using parameters that govern their dynamics [54]. In another model exploring marital relationships, researchers implemented "emotion" states and parameters that dictate how married couples influence each other's emotions [55]. Yet, these models often remain abstract, with behavior not directly represented (e.g., Buder [54] interprets a coupled dynamical system *as* interaction).

With the model we present here, we pursue a theoretical framework in which we specify some basic properties of agents' interaction and then examine their coupled dynamics. Importantly, our approach facilitates the integration of the agents' "task" in a manner that is flexibly defined, enabling it to account for the emergence of synchrony and complementarity in different contexts. We identify a set of components comprising an interaction, including the task context, the behaviors of two agents, and how they influence each other. The parameters of these components can be specified manually and simulated under various conditions, resulting in a comprehensive exploration of how and why two simulated agents become aligned or complementary according to specific principles. Importantly, this computational investigation can inform how the task itself may govern the emergence of these behavioral processes. This framework contrasts with prior computational efforts that have focused on model development in a way that centers on the task. In prior coupled-oscillator models, for example, there typically exists a conception of two or more systems that are directly coupled in an interaction, shaping their mutual dynamics as a function of time. We don't discount this coupling as a key feature of interaction, but our model centers the task as a potential source of constraint on coupled dynamics. A theoretical implication of this focus on the task is that interaction partners may not need to retain all details of an interaction to support their joint performance. Instead, the task environment can influence the agents' dynamics without the agents incurring high cognitive demands.

Beyond devising a computational framework to explore alignment and complementarity, a second motivation of this approach is to bridge the model with human data from real tasks eventually. This integration with human data is facilitated by having a model whose inner-workings are relatively easy to understand and modify. The model can be employed to predict how various interactive tasks with human participants yield distinct signatures of alignment and complementarity. After presenting our modeling framework, we illustrate two ways in which the model can capture data collected in experimental human tasks. In our concluding discussion, we also address how bridging the model to human data can be explored in future model development.

## Modeling Framework

Our initial approach is to draw from discrete dynamical systems models that permit direct specification of parameters and behaviors [56]. We model a single behavior for two interacting partners, and represent these behaviors as a two-dimensional vector that is updated iteratively over time (one dimension per simulated agent). We take their behavior as numeric descriptions, scalars, which could describe many quantitatively measured behaviors in real interaction (e.g., eye movements, body motion, etc.). Specifically, the scalar number could represent the attentiveness of an agent at a specific time point, or the rate of speaking or arm movements. The model is deliberately abstract in this regard, allowing for its application across a wide variety of operationalizations in human tasks.

Taking $B$ to be this vector ($b_{\text{Person 1}}$, $b_{\text{Person 2}}$) at time $t$, we have:

$$B(t) = C \bullet I \bullet B(t-1) + U(-.5, .5) - \alpha B(t-1)$$

The behavior of two people $B(t)$ is a function of a given context $C$ and power of influence $I$. We take this $C$ parameter to be a "context matrix," because it transforms two-person behaviors based on a prior time step ($t-1$), with added noise ($U$). In this iteration, $I$ is set as constant ($I$=1), which represents 100% receptivity of two people towards all signals. $U$ is set in range (-.5, .5), which represents a uniform source of noise per turn, from -.5 to .5. The final subtractive term is a decay term with scalar α, generally a number between 0 and 1, which we specify as 0.1. This decay term proportionally reduces the behavioral signals by 10% to ensure that the model does not saturate and veer towards positive or negative infinity.

The task context $C$ is numerically described as a 2x2 matrix whose values can be 0, 1, or -1, representing inactive, active, or inhibitory constraints on communication. For example, during a presentation, the speaker is actively speaking, while the audience is inactive in speaking. If an audience member has a question in mind but inhibits the act of asking questions immediately to maintain the presentation flow, they are practicing inhibition. This $C$ matrix specifies how much individuals influence each other, how much they are autocorrelated (influence themselves), and can vary across *specific* tasks that individuals face. The 2x2 matrix consisting of elements $S_1$, $O_1$, $O_2$ and $S_2$ can be interpreted in the following way:

$C = ($ $S_1$ how much *Person 1* influences self, $O_1$ how much *Person 2* influences *Person 1*;

$O_2$ how much *Person 1* influences *Person 2,* $S_2$ how much *Person 2* influences self $)$

Fig. 1 further illustrates the interpersonal influences constructed in $C$. A consequence of our modeling approach is that the configuration of values in the context matrix $C$ represents task constraints and reflects different interaction types. For example, if $C$ is "null" with $C = (0, 0; 0, 0)$, the model generates agent behaviors that are random with no meaningful structure or interaction, and $C = (1, 1; 1, 1)$ represents a perfectly synchronizing interaction, as when two people are singing a song in unison. When $C = (1, 0; 1, 0)$ or $(0, 1; 0, 1)$, there is a role asymmetry, with one person actively influencing the other's (and their own) behavior. This would be relevant to educational scenarios, such as lectures, where one partner primarily drives the interaction. We can also implement *inhibitory* constraints, such as $(1, 0; 1, -1)$, which could again represent a lecture scenario, but this time, the listener has a question in mind and inhibits from asking the question so that the speaker can maintain the communication flow (see Fig. 2.iii and Fig. 4 for a zoomed-in version).

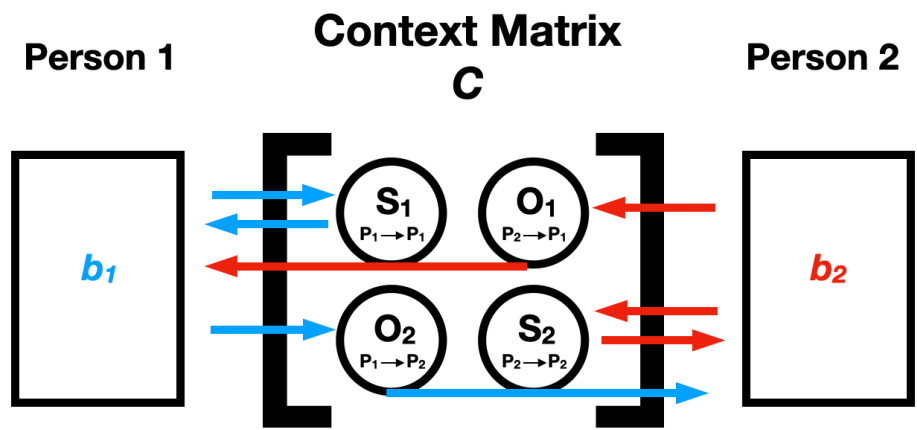

Fig. 1: How context matrix $C$ mediates interactions between two agents (blue, red are Person 1 and Person 2) and their resulting behaviors ($b_1$ and $b_2$ for Person 1 and 2 respectively). $S_1$ represents how much *Person 1* influences self and $S_2$ how much *Person 2* influences self. $O_1$ represents how much *Person 2* influences *Person 1* and $O_2$ how much *Person 1* influences *Person 2*. Arrows represent the direction of influence, and the color of the arrows corresponds to the person initiating the action (blue for Person 1-initiated action and red for Person 2-initiated action).

## Simulation Methods

We implemented a simulation under this framework to explore the impact of $C$ on the behavior of the two coupled agents. We aimed to investigate *all* potential variations of the $C$ matrix. Exploring all possible combinations of the parameters of $C$ will allow us to isolate what governs the emergence of synchrony or complementarity in this model.

***Independent variable: C***. Our model of a two-agent (Person 1 and Person 2) system permitted us to codify dyadic interactions using context matrix $C$, which represents four possible combinations of influence: (i) Person 1 to Person 1, (ii) Person 2 to Person 1, (iii) Person 1 to Person 2, and (iv) Person 2 to Person 2. Recall that each of these four directions of influence has three possible states – active, inactive, and inhibitory – represented by 1, 0, and -1. With these numeric codes, we are able to characterize different tasks or contexts in human interaction using the context matrix $C$.

***Dependent variable: r***. To account for the relational nature of behaviors, *Pearson's r*, or correlation coefficient, is often used as the dependent variable when investigating interpersonal synchrony [45]. In addition to being one of the most common metrics used to describe interpersonal synchrony, $r$ also allows clear and familiar interpretation of the correlation between two signals. If $r$ is strongly positive, it would indicate two agents that are synchronizing in whatever behavior they exhibited in the "conversation"; if $r$ is strongly negative, it would suggest a kind of complementarity – doing the opposing behavior of the other (akin to taking turns in speech, or taking on separate complementary roles in an interaction). Since $r$ is a statistic describing how two agents correlate with each other in interactions, the meaning of "complementarity" could have multiple interpretations in different contexts. For example, it could represent the "division of labor" in collaboration, where agents may engage in distinct, concurrent behaviors throughout the task. Complementarity could also signify "turn-taking" in conversations, where agents engage in distinct behaviors simultaneously (such as one being silent while the other talks), which alternate between agents over time.

In the simulations discussed below, we approach interpersonal synchrony and complementarity by directly modeling the correlation coefficient ($r$) between two agents. We use the simulated correlation coefficient ($r$) as the dependent variable, we then perform regression analysis to further evaluate and interpret this computational quantification of interpersonal synchrony. Finally, we apply the model to simulate human data from prior studies [56] using a different measure of interpersonal synchrony – cross-correlation functions. We show that our model's $\boldsymbol{C}$ matrix produces distinct cross-correlation functions that resemble the patterns of synchrony and complementarity documented in these studies. By directly modeling the statistics that other human studies use to quantify synchrony, we bridge computational simulation and empirical quantification as methodologies for exploring behavioral dynamics.

***Simulation procedure.*** Using the model presented above, we obtained simulation data with the following steps:

**Step 1:** We obtained all possible combinations of $\boldsymbol{C}$. With 4 parameters and 3 possible values (-1, 0, 1) for each, there are $3^4 = 81$ combinations of the C matrix. These combinations represent all possible interactive scenarios involving the two agents. We generated all such permutations of $C$ in rounds of our simulation.

**Step 2:** We obtained simulation data and calculated the dependent variable $r$. We ran 100 simulations under random starting conditions for each of the 81 possible $C$ matrices, resulting in 8,100 simulations. For each simulation, one of the agents is chosen to "start the conversation," and displays its behavior (element of $B$) to the other; the model then operates with its matrix $C$, modifies agent behaviors, and the next agent's turn commences. This is done for 500 turns for each simulation, and we record each sequence of behavior separately. These behaviors represent two time series, and after each such simulated "conversation," the correlation coefficient $r$ between these behavioral sequences is recorded. As noted above, this coefficient $r$ represents relative synchrony or complementarity and is utilized as the dependent variable for later analysis.

**Step 3:** We then conducted a regression analysis, using $C$ to predict $r$. This regression model takes values in $C$ to predict the resulting $r$ across the 8,100 simulated interactions. Specifically, we performed multicategorical analysis and built various regression models that include different subsets of $C$ in order to isolate the strongest contributors to determining $r$. Multicategorical analysis is commonly used to unpack all aspects/categories of a variable with multiple features, and regression models are used to understand how each aspect of the independent variable explains/predicts the dependent variable. In this case, components of $C$ include $S_1$, $O_1$, $O_2$, and $S_2$. These regressions are simple linear models that account for the simulation and help us understand how different context matrices $C$, or different combinations of active, inactive, and inhibitory states of communication influence the relative dynamics between the two agents. The regression analysis helps us understand what drives this theoretical model, offering answers to some key questions: Is it possible for a simple regression model to capture complex interpersonal dynamics from the task matrix $C$? How many parameters and what kinds do we need to make this prediction? For example, is it possible that the self-related parts of $C$ – how agents influence themselves – are sufficient to predict $r$ across simulated interactions?

All code and data necessary to reproduce Steps 1 to 3 are found at the following GitHub repository: [https://github.com/miaoqy0729/sim-syn-sims](https://github.com/miaoqy0729/sim-syn-sims).

## Results

### *Trends in C for predicting r*

As noted, we performed 100 simulations for each of the 81 possible configurations of the task context $C$, involving active, inactive, or inhibitory states for each of the 4 possible directional influences within and across interacting partners. Illustrations of the output of model simulations are shown in Fig. 2, including when agents have no or weak interdependence (Fig. 2.i and Fig. 2.iv), when two agents perform optimal synchrony and yield high interdependence (Fig. 2.ii), and when one agent leads the interaction and the dyad yields complementary interdependence (Fig. 2.iii; see also Fig. 4).

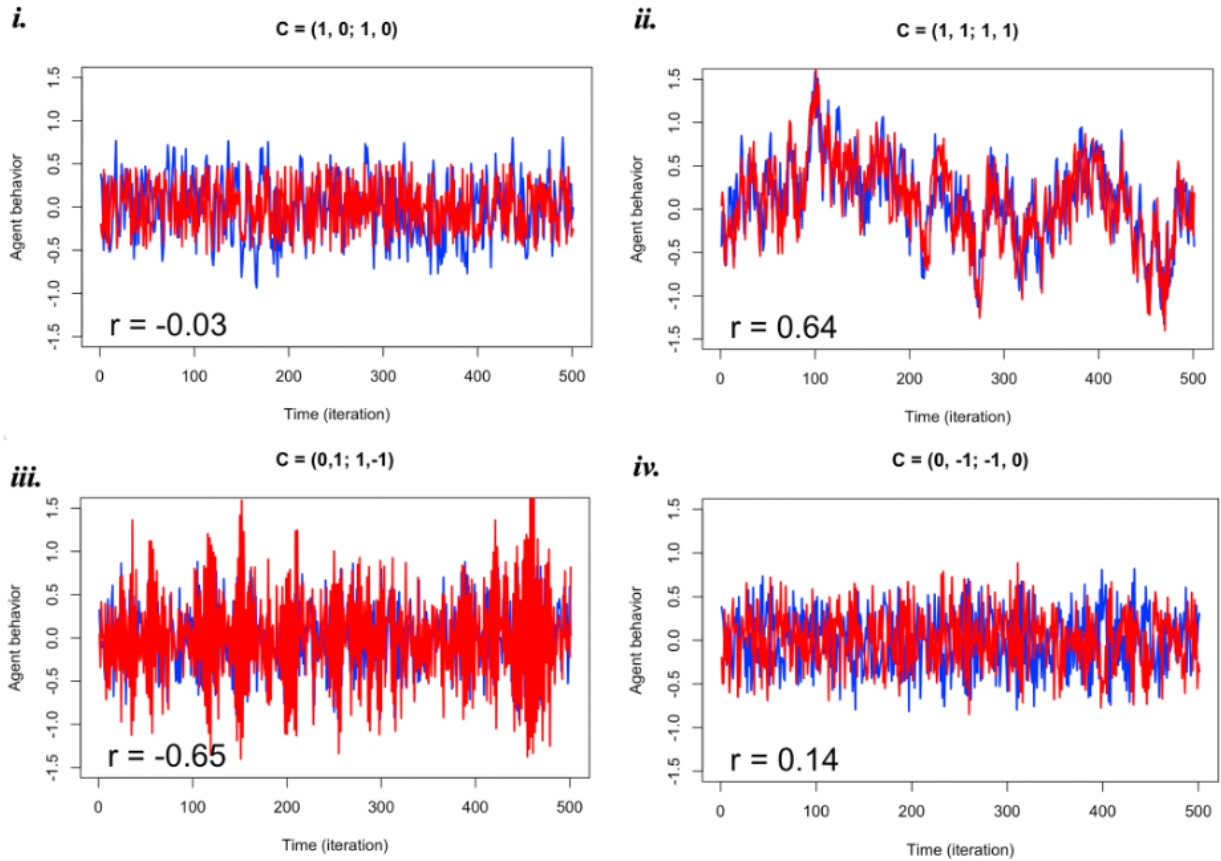

Fig. 2: Time series of two agents (blue, red are $b_{\text{Person 1}}$ and $b_{\text{Person 2}}$) under different context matrices. i. When Person 1 is active but Person 2 is inactive, they have no interdependence. ii. Total coordination between two agents (such as singing a song together) produces close synchrony. iii. In a leading/following scenario (such as teaching), blue precedes red and correlates in their simulated behaviors. (Fig. 4 shows a zoomed in segment of these time series). iv. With inhibitory parameters, a weaker correlation (calculated using 0-lag cross-correlation) is exhibited by the simulated dyad, and their behavior fluctuates between states.

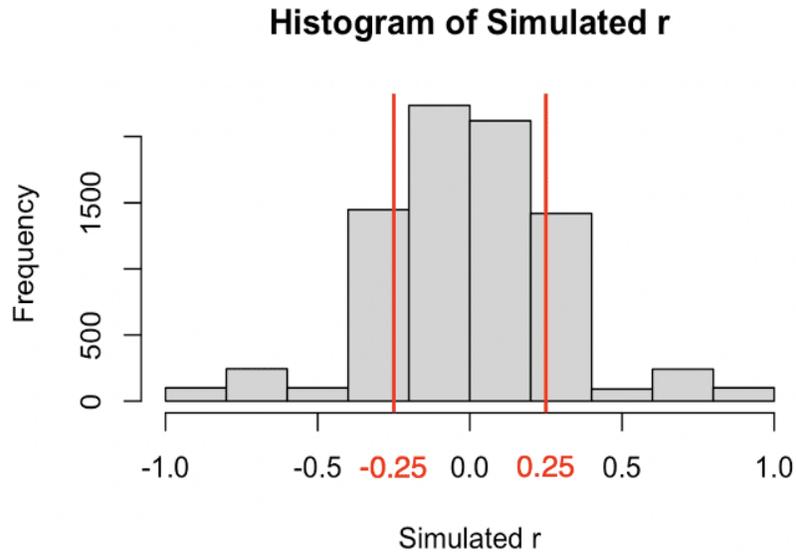

Fig. 3: Frequency distribution of correlation coefficient ($r$) obtained from the 8,100 simulations. Red lines mark the thresholds (r < -0.25 for complementarity and r > 0.25 for synchrony) for statistical tests.

The model exhibited a range of behaviors. The distribution of $r$ across the 8,100 simulated "interactions" is shown in Fig. 3. Sometimes agents show perfect synchrony, with $r \sim 1$; agents can also exhibit extreme complementarity, $r \sim -1$. Since we were interested in how the context matrix may determine these behaviors, we inspected what characteristics of $\boldsymbol{C}$ seemed to accompany extreme values in $r$.

We considered the negative correlations in the simulated results, which represent complementarity – complementary actions performed at the same time when engaging in an interpersonal activity. For example, Fig. 4 illustrates a leading/following scenario where Person 1's behavior (blue) precedes Person 2's behavior (red), while being highly complementary to each other ($r$ = -0.65).

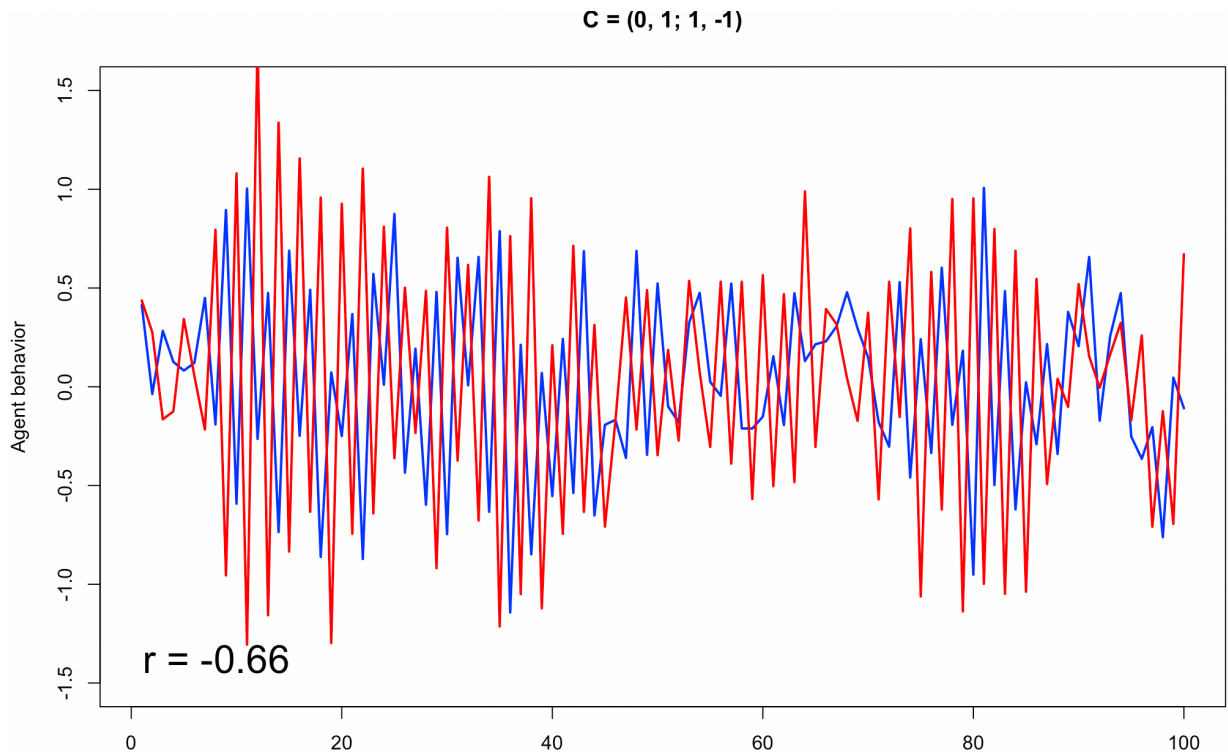

Fig. 4: Zoomed in time series of Fig. 2.iii. In a leading/following scenario (such as teaching), blue precedes red and correlates in their simulated behaviors.

Across the 8,100 simulations, we examined when models exhibited a strong negative correlation in agent behavior akin to complementarity. We defined the threshold as $r < -0.25$, as this captures the left side of the observed distribution of $r$ outside of its mean (see Fig. 3). In these cases, the task matrix ($C$) contained at least one $-1$ parameter 99.8% of the time. To test for statistical significance of this observation, we conducted a chi-square test on uniform probability, yielding a significant result ($\chi^2 = 1317.00$, $p < .00001$). We also conducted an additional test using probabilities of the corresponding positive correlations in the simulations as baseline (using a $r > 0.25$ threshold). The observed rate of negative task parameters was 83.3%, which is significantly different from that in the negative-$r$ threshold ($\chi^2 = 259.98$, $p < .00001$). This suggests that in order for there to be a robust complementary pattern (low, negative $r$), the combination of constraints in $C$ must include at least one negative value almost all the time. It is important to note that the structural features of the model may create certain consequences. For example, constraints on some models might result in patterns like "turn-taking" simply by virtue of their underlying mathematical formulation, such as in predator-prey population levels modeled with difference equations [56]. A similar constraint may be present in our formulation as well. Nevertheless, our result supports the broader hypothesis, discussed in the introduction, that inhibition is a critical aspect of cognitive processing central to successful coordination. In fact, there are almost no conditions of $C$ that can yield complementarity dynamics without inducing inhibitory constraints. To delve into the subtleties of this result, we performed further analyses using multiple regression.

### *Predicting r from C with multiple regression*

The inhibitory effect we documented suggests that the model's formulation produces constraints that can predictably result in specific values of $r$. In the most pronounced case we observed, extreme negative correlations appeared to require inhibition. Consequently, we sought to understand how specific values of $C$ contribute to specific behavioral patterns, such as which values predicted synchrony and which predicted complementarity. To answer these questions, we performed a multicategorical analysis on the simulated data described earlier.

Specifically, all four parameters of the $C$ matrix ($S_1$, $O_1$, $O_2$, $S_2$) were dummy coded with value 0 as the reference group, resulting in 8 parameters ($S\_1p$, $S\_1n$, $O\_1p$, $O\_1n$, $O\_2p$, $O\_2n$, $S\_2p$, $S\_2n$) representing each C matrix ("p" and "n" stand for "positive" and "negative"). With the positive and negative values specified for each of the four parameters, regression analysis can reveal whether the valence of three different states of communication plays impacts the communication flow. In addition, to better explore the components that account for correlations of dyadic behavior in the simulated data, linear interactions across all possible dummy-coded parts of the model are calculated.

We focus on three regression models containing different predictors that account for the dyadic correlation: (1) a model that includes the main effects of the 8 dummy coded parameters in the C matrix (S_1p, S_1n, O_1p, O_1n, O_2p, O_2n, S_2p, S_2n), (2) a model that includes the possible interactions among different parameters (8*6 = 48 interactions), and (3) an overall model consisting of all predictors (8 + 48 = 56 predictors).

To evaluate the amount of variability in the simulated data the regression model could explain, we considered the $R^2$ values of the three regression models. The main effects model had an $R^2$ value of 0.101, accounting for 10% of variability in the dependent variable $r$. The interaction model had an $R^2$ value of 0.945, thus accounting for 94.5% of the variation in $r$. The overall model with 56 predictors had an identical $R^2$ value as the interaction model by itself. These $R^2$ values suggest that when predicting synchrony between two agents, optimal results could be achieved solely by the interaction parameters. As illustrated in Fig. 5, the regression model with only main effect terms failed to yield fine-grained predictions, whereas the model with interaction terms predicted $r$ almost perfectly. In other words, the key features of how the task constrains two agents *relative to each other* can alone predict the agents' coordination ($r$) almost perfectly.

We further examined whether an $R^2$ value greater than 0.9 can be obtained using subsets of the predictors of the overall model, instead of all 56. We found that more than 90% of variability could be accounted for with fewer than half of the 56 predictors. For example, the $R^2$ of the model (4) that accounts for how Person 1 influences themselves (S_1p, S_1n) and Person 2 (O_2p, O_2n) is 0.91 (see Table 1). And the $R^2$ of the model (5) that accounts for how Person 1 influences themselves (S_1p, S_1n) and how Person 2 influences themselves (S_2p, S_2n) is 0.94. This means more than 90% of the variability in the dyadic system could be estimated by focusing on a specific part of the dyadic system – either focusing on the initiator of the time series or on the agents' auto-correlation with themselves. We explain further our rationale for selecting the predictors of models (4) and (5) below, and evaluate their efficiency.

To determine the optimal model that accounts for an adequate amount of variability while using a minimal number of predictors [57,58], we calculated the two popular model selection criteria – Akaike information criterion (AIC) and Bayesian information criterion (BIC) – for each regression model. In general, the *lower* AIC and BIC for a given model, the more effective the model is at balancing complexity and variance captured of the observed data – namely, a more efficient model.

| Model | Number of Parameters | $R^2$ | Adjusted-$R^2$ | AIC | BIC |
|---|---|---|---|---|---|
| (1) Main effects | 8 | 0.101 | 0.1001 | 2851.535 | 2921.532 |
| (2) Interactions | 48 | 0.9451 | 0.9449 | -19743.45 | -19505.47 |
| (3) Overall | 56 | 0.9451 | 0.9449 | -19743.45 | -19505.47 |
| (4) Initiator-focused | 28 | 0.9084 | 0.9081 | -15606.81 | -15396.82 |
| (5) Self-influence | 28 | 0.9405 | 0.9403 | -19102.45 | -18892.46 |

Table 1: Information of various regression models. Model (1) includes all main effect terms (S_1p, S_1n, O_1p, O_1n, O_2p, O_2n, S_2p, S_2n) as regressors. Model (2) includes all the interaction terms (for example, S_1p * (O_1p + O_2p + O_2p + O_1n + O_2n + S_2n) contains all interaction terms for S_1p) for each of the 8 main effect terms. Model (3) includes all of the regressors in Model (1) and Model (2). Model (4) focuses on Person 1 as the initiator, and include regressors on how Person 1 self-influences and how Person 1 influences Person 2. Model (5) includes Person 1's self-influence and Person 2's self-influence. In other words, how agents auto-correlate with themselves, or how Person 1's past behavior predicts Person 1's future behavior.

The lowest AIC and BIC values came from the models with interactions only or with the full model. These two models had the same information score (AIC = -19743.45, BIC = -19505.47, see Table 1). Because the interaction-only model has fewer terms compared to the overall model, we can regard it as more efficient. The least effective model was the one with main effect terms on their own, which had the highest score on AIC and BIC.

In addition to testing models based on their statistical properties (namely, main effects or interactions), we used subsets of the 56 predictors to devise models to answer specific questions of interests: How much variability in the overall distribution can we account for by focusing on the influences of the initiator? Or by focusing on the self-influence within each agent?

To answer the first question, we focused on Person 1 as the action initiator in Model (4), which involves $S_1$ (how much *Person 1* influences self) and $O_2$ (how much *Person 1* influences *Person 2*). The parameters in Model (4) include s_1p, s_1n, o_2p, o_2n and all the interaction terms involved. To answer the second question about the role of self-influence, we considered model (5) focusing on $S_1$ (how much *Person 1* influences self) and $S_2$ (how much *Person 2* influences self). The parameters in Model (5) include s_1p, s_1n, s_2n, s_2n and all their relevant interaction terms. With an equal number of parameters, the self-influence model (see Table 1 for model 5) has noticeably better AIC and BIC results than the initiator-focused model (see Table 1 for model 4).

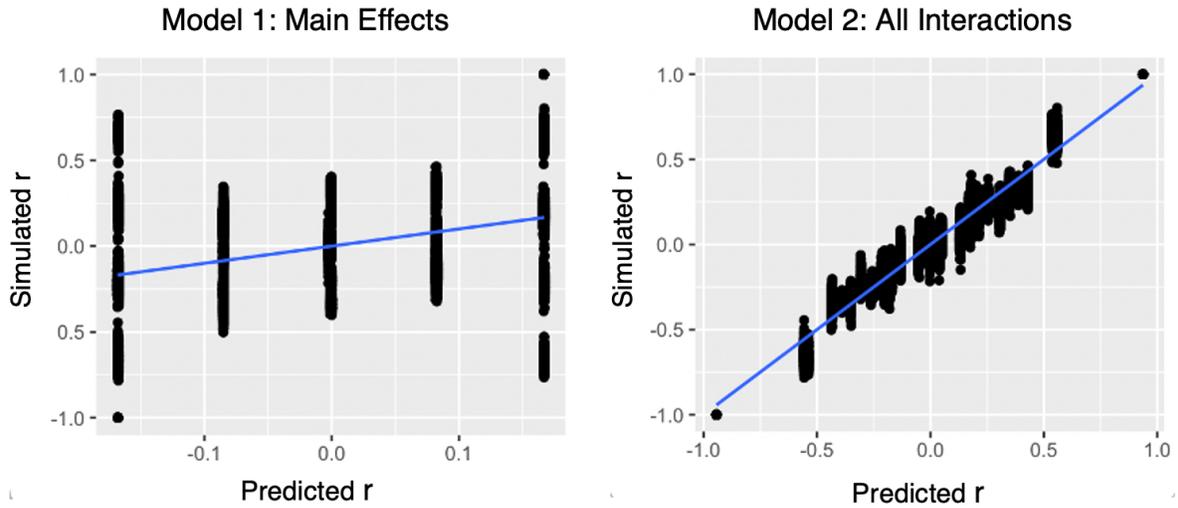

Fig. 5: Predicted results of Model 1 (main effects model) and Model 2 (all interactions model) (x-axis) and $r$ value between two agents (y-axis). Model 1 is not able to make fine-grained predictions, while Model 2 predicts $r$ almost perfectly. Further context for Model 1's failure to capture the trends in $r$ comes from the AIC and BIC values in Table 1: they indicate that all models with interactions are more efficient and should be selected over the main effects model. In particular, the AIC and BIC both suggest that Model 2 is a more efficient model in terms of complexity and data fit.

Despite being a highly simplified exploration of these effects, the framework offers a foundation for understanding how behavioral dynamics between partners may be accounted for with task constraints. These regression results suggest that relatively efficient information from a task (the task matrix $\boldsymbol{C}$) can be used to predict how interlocutors in the simulation will synchronize under that task.

**Comparison to Human Tasks**

So far, we have demonstrated that our model can capture essential features of how the task can mediate behavior among agents, and that the task alone can predict the aggregate dynamics exhibited by the agents. In this section, we show how this framework could bridge to properties of human experimental data, providing two specific illustrations.

In the first illustration, we use cross-correlation in a manner related to prior interpersonal dynamics [32,59], and show that the model's $C$ matrix can moderate this signal in interesting ways. Results indicate that while the $C$ matrix generates the common observation of a peaked cross-correlation function, this function is modulated depending on the task setup of $C$.

In the second illustration, we show that turn-taking dynamics can also be captured with this model, with $C$ again playing a modulatory role [60]. As Stivers and colleagues [60] have shown, turn transitions can be quite rapid, across different cultures. Together, these illustrations show how our simple framework can produce nuanced patterns resembling those observed in human tasks. They also reveal instances where the simulation *diverges* from what would be expected or possible in the human scenarios.

### *Cross-Correlation Function*

In the first case, we examined several example $C$ matrices that demonstrate various patterns including synchrony, anti-synchrony, and leading and following, all of which have been observed in human dynamic data [38,47,53]. When neither agent is coupled to the other ($O_1$, $O_2$ = 0), the cross-correlation function appears flat (Fig. 6, green). However, a specific $C$ matrix that induces leading and following, in that the cross-correlation function is shifted, further emphasizes the potential role of inhibition. In situations where one agent follows the other ($O_1$ = 1) and the task "inhibits" their own autocorrelation ($S_1$ = -1), the result is a shifted cross-correlation function towards one of the agents leading (Fig. 6, purple).

In examining this concordance with human data, we discovered that some $C$ matrices produce implausible cross-correlation functions. Rather than being a weakness, this can be seen as a strength of the model: it suggests that this modeling framework could be used to evaluate whether certain $C$ matrices are cognitively plausible at all. For example, if both agents inhibit their own behavior while trying to mimic the other, their mutual "attractor" becomes unstable, because the agents are simultaneously seeking "mimicry" (the same behavior) and the $C$ matrix restricts this possibility. The result is an erratic cross-correlation function (Fig. 6, right). We revisit this issue in the conclusion, arguing that the modeling framework need not produce coherent human-like responses in all cases; when it does not, it may be suggestive of unstable task contexts.

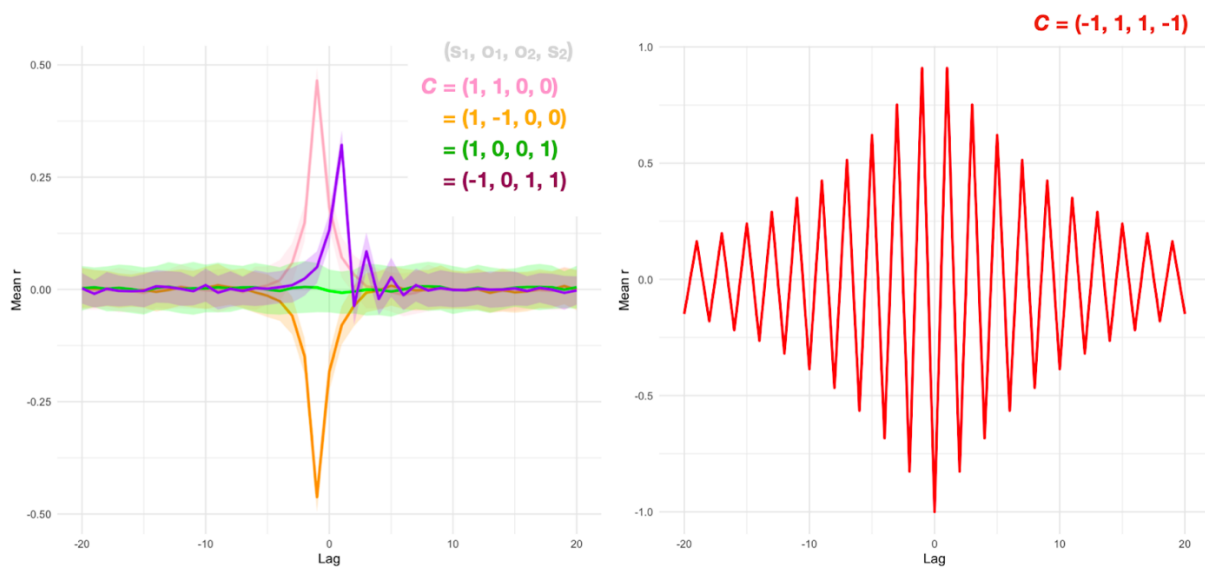

Fig. 6. Mapping human task dynamics with simulation data. Left: Average cross-correlation functions for different $C$'s (ribbon width = SD). $C$ parameters that promote other-influence yields positive correlations, but this profile is shifted when one agent "inhibits" its own states but follows another. Right: Some $C$ configurations are unstable, such as two systems avoiding their own behavior while seeking to "mimic"; the system cannot stabilize because it at once seeks to converge and avoid similar behaviors.

We also examined the behavioral time series from both agents in $\mathbf{B}(t)$ and assessed when they "take turns." We did this by thresholding the state variables of these systems, taking $b_1(t) >$ $\mu_1$ and assessing the closest point in time that $b_2(t)$ showed an "on" state. We recorded these "lags" between the times when the two agents became more activated, drawing an analogy to the human data in Stivers et al.[60], who showed rapid transitions during polar (yes-no) questions. Humans tend to respond so quickly that the highest maximum is near 0. The researchers also found some subtle cultural influences that altered this turn-taking lag (see their Fig. 1). When we performed a similar analysis with our simulated agent data, we found a qualitative resemblance to the lags described in Stivers et al. [60]. Moreover, these lags do seem to be influenced by *C*, showing that subtle impacts of the task setup may slightly alter the structure of turns. Additionally, and somewhat speculatively, these subtle impacts of *C* may hint at methods for conceptualizing cross-cultural differences. In this context, culturally specific patterns of turn-taking can be regarded as subtle changes to how participants in interactions perceive the *C* that underlies their interaction (see discussion of subtle cross-cultural variations in [60]. This lag result is shown in Fig. 7 according to the same conditions shown with cross-correlation, revealing the slight modulation of turn-taking patterns by the task matrix.

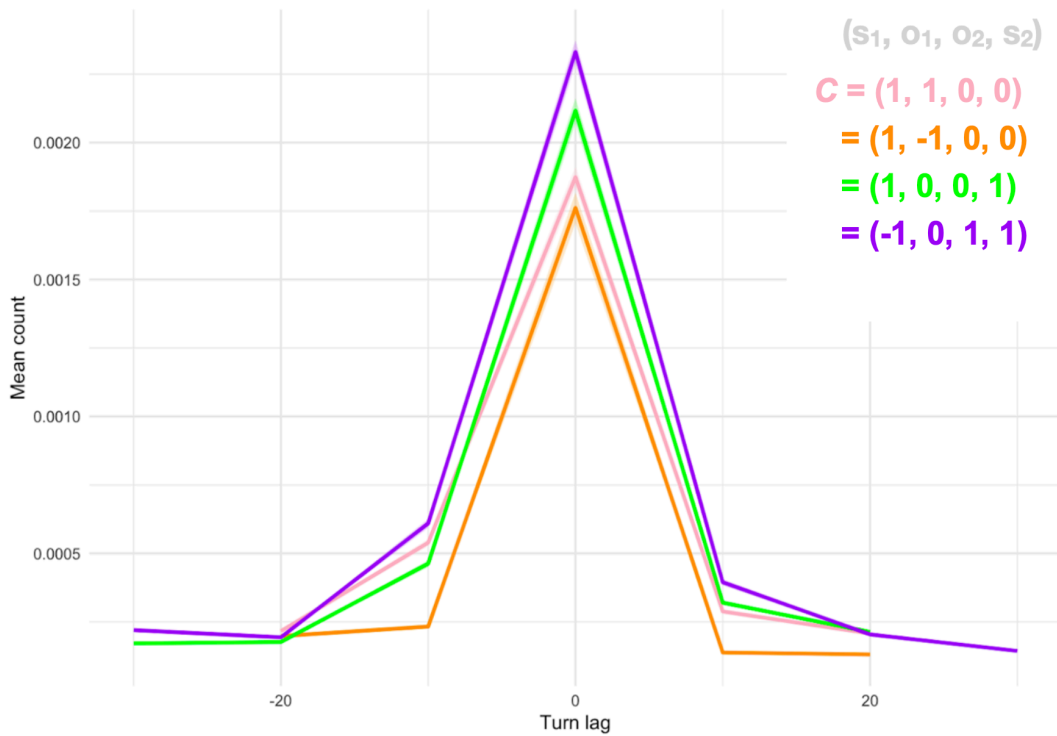

Fig. 7. Mapping gaps in human conversational turn-taking patterns from Stivers et al. [60] with simulated data. Different colored lines represent various ***C*** matrix inputs and the generated patterns over different time lags (from $t_{-20}$ to $t_{20}$).

## Discussion and Conclusion

Dyadic collaboration has been widely studied empirically, but computational models in this domain have focused on accounting for motor coordination, such as joint actions [35,44-47], or testing specific cognitive mechanisms, such as action prediction and signaling [39]. We introduced a simple modeling framework, relevant to both joint action and conversation, that simulates the emergence of the dynamics of partners' behavior across varied collaborative task contexts. The framework allows us to transparently couple task parameters and behavioral dynamics: It allows us to implement and test that changes in the task constraints, through the context matrix ***C***, lead to different stable patterns of relative behavior, expressed with *r* and other metrics of interpersonal synchrony (e.g., cross-correlation function). Results from various matrices can, in combination, simulate dyadic interaction from the onset of interaction to emerging synchrony and complementarity.

We evaluated variation in the context matrix $C$ to compare how different inducements during simulated communication (active, inactive, or inhibitive) would lead to different levels of alignment between task partners. Our analysis suggests that in order for complementarity to emerge in the simulation (such as in turn-taking scenarios), at least one of the two agents needs to be actively inhibited by the task. This inhibition is linked to self-restraint from action.

Although this result might be an inherent outcome of the model's specification, it provides an essential insight revealing that our framework is capable of probing the fundamental structural ingredients of interactions. It is important to recognize, however, that this initial finding allows only limited interpretation because our simple model lacks the complex and subtle distinctions between different kinds of inhibition, such as the exogenous inhibition introduced by tasks, and the endogenous inhibition involved in the internal cognitive process.

Despite this limitation, our discovery regarding the significance of inhibitory parameters in producing clear complementary dynamics is consistent with other research findings that highlight the role of executive processes in communication and collaboration [14,61]. For example, there is evidence that individual differences in inhibitory control significantly predicted people's ability to inhibit their own perspective and consider discrepancies in shared information during language processing [14]. Conceptually, these findings are consistent with wider theoretical proposals of synergistic self-organization, a concept used to explain the behavior of complex dynamic systems. Under this framework, cognition and behavior in a multi-agent environment exhibit distinct and partly stable patterns of coordination emerging from constraints across people and the task [8,56,62]. We would argue that a perspective rooted in synergy elegantly frames the initial modeling endeavors reported here.

In this general and simple framework, we have included only a few essential components of human interactions, such as agent behaviors and context. This simplicity provides flexibility for future iterations of the model that may focus on various aspects of human interactions, such as different potential loci for where inhibition resides (whether cognitive control, or task control). By mapping the human data collected in Stivers et al. [60], we demonstrated how this simple framework can be applied to specific scenarios like turn-taking patterns of polar question answering in daily conversations. Future iterations of the model can increase in complexity to investigate the nuances in human interactions that are abstracted away in this first illustration, so that model and theory may serve together as a framework for exploring key characteristics of interpersonal processes through a computational lens.

Though promising as a preliminary framework, our model has various limitations in addition to the simplified nature described above. For example, we currently represent behavior as scalar signals instead of specific behavioral metrics. This is a common limitation shared among computational models that aim to represent interpersonal interactions [35,54]. To move beyond these initial steps, we identify three possible future directions to explore how computational endeavors can achieve increased internal and external validity in computationally characterizing human social interaction and cooperation.

### *Information receptivity*

The speaker's power of influence and the listener's level of information receptivity are described by the $I$ parameter in the model. The current model assumes $I = 1$ in order to exhaustively explore the effects of context matrix $C$. However, in real life, it seems very unlikely that persons attend to every meaning, and few if any individuals can exert maximum power of influence on another person's behavior and thoughts. This limitation is not restricted to the model we derived. In fact, there is a lack of modeling endeavors looking into the dynamics of influence during social interaction and cooperation.

The $I$ parameter in our model can be adapted to capture the dynamics of influence between interlocutors. To achieve this, we plan to build on the theoretical construct of communicative potential (CP) – an attribute that describes the level of receptivity of the information recipient, or the potential that a certain piece of information will be communicated [63]. For example, we may transform $I$ from a scalar to a function with various parameters that account for factors involving information receptivity, such as the listener's level of interest, complexity of the topic, and psycho-physiological states for speaker and speaker at a specific time point.

### Modality of behavior

The current model uses a generated number to represent one behavior, such as eye gaze or speaking. These generated numbers are extracted from a uniform distribution with random deviations, which represents a single modality of behavior in real life. However, the interpersonal processes in real life are multimodal in nature [64]. When we interact with people, multiple modalities of perceptual signals take place concurrently – including vocal, facial, gestural signals – along with neurocognitive activities. Different behaviors can have complementary roles – there could be complementarity in one dimension (e.g., eye movements) but still high alignment in language use. This limitation is not restricted to the model we derived.

In fact, most dynamical models focus on one aspect of interaction, such as conversational turn-taking [54] and motor control in spatial movements [35]. However, dynamical models don't typically synthesize these modalities in a single model to capture the multimodal nature of human interaction.

We plan to update the *B* parameter in our model from a two-dimensional vector with one dimension per simulated agent with one behavior (such as eye movement), to a three-dimensional vector with two or more behaviors per agent (such as both eye movement and conversation). This three-dimensional vector will be able to describe multiple behavioral components concurrently to mimic real-life interactions.

### *Behavioral consistency over time*

The current model includes the power of influence, context of the interaction, and behavior represented by a generated number as the initial parameters, then regurgitates the behavior at previous time points to generate behavior of the current time point. In other words, the interactional process that happens at an early time point (e.g., $t = 1$) is identical to that at a later time point (e.g., $t = 100$). While permitting statistical testing of critical cognitive processes in interaction, such as inhibition, this stable interactive dynamic is not an accurate representation of interpersonal processes in real life. Interactions have complex structures, and are not given to one single stream of behavioral patterning [65].

One possible solution is the segmentation of interaction — transforming a continuous interaction into segments with distinct features. For example, a recent paper by Microsoft Semantic Machines shows how dialogues can be segmented and represented as flow charts [66]. This relates to statistical methods in machine learning, such as the automatic distillation of structure (ADIOS) algorithm [67], and other discourse analytic methods, such as dialogic resonance [68]. Other work in cognition and communication suggests that distilling structural patterns of interaction is important for understanding complex combinations of emergent patterns that include both synchrony and complementarity. For example, Fusaroli and Tylén [69] found that people are most effective in an interactive problem when they form more complex patterns of communication, referred to as interactive "synergy" (see also [32]), and not merely synchrony. Our model may be able to capture some trends in conversation, such as emerging complementarity, but because their dynamics are sustained in the same way across simulated conversations, they cannot easily model abrupt changes in behavior. Integrating these ideas into the model would diversify the possible patterns emerging in dyadic interaction. For example, an interaction could show distinct leading and following patterns as conversation topics shift, or as goals shift in the interaction. Such sequential structuring of interaction dynamics would allow greater ecological validity in our computational investigations of interpersonal dynamics.

The initial results and versatile extensions of this simple computational framework could further accommodate the variability and complexity of social interactions in human cognition and behavior. Interpersonal communication is a complicated multiscale process involving coordination in conversational turn-taking [70], multimodal behavior, including gaze and gestures [64], and even correspondence in neurocognitive activity [71,72]. Unpacking this multiscale nature of communication requires computational models that are highly flexible, extendable, and interpretable. Devising such a model for dyadic interactions can potentially further inspire the development of computational models for teams and role emergence in groups [73,74]. In conjunction with computational modeling, bridging to the efforts of other methodologies and disciplines that investigate social interactions – including conversation analysis, experimental social psychology, and social neuroscience – could help to address open queries about the multiscale nature of human communication.

**Data Availability**

Code used in this study is available on GitHub at https://github.com/miaoqy0729/sim-syn-sims.

**Acknowledgement**

This project is funded by National Science Foundation grant #2120932 Identifying multimodal signatures of coordination to understand joint performance in diverse tasks.

We declare no competing interests.

**Author Contributions**

A.G., G.M. and R.D. planned the research. G.M. and R.D. designed and programmed the simulation. G.M. programmed the regression analysis. G.M., A.G. and R.D. wrote and revised the paper repeatedly.